# An Assessment Tool for Academic Research Managers in the Third World

Delbianco, Fernando
Departamento de Economía, Universidad Nacional del Sur – INMABB-CONICET
fernando.delbianco@uns.edu.ar, ORCID: 0000-0002-1560-2587
Phone/Fax +54 (0291) 459 5138 (int. 2712)

Fioriti, Andrés
Departamento de Economía, Universidad Nacional del Sur – INMABB-CONICET
andres.fioriti@uns.edu.ar, ORCID: 0000-0002-6771-0643

Tohmé, Fernando
Departamento de Economía, Universidad Nacional del Sur – INMABB-CONICET
ftohme@criba.edu.ar, ORCID: 0000-0003-2988-4519

**Abstract**

The academic evaluation of the publication record of researchers is relevant for identifying talented candidates for promotion and funding. A key tool for this is the use of the indexes provided by *Web of Science* and *SCOPUS*, costly databases that sometimes exceed the possibilities of academic institutions in many parts of the world. We show here how the data in one of the bases can be used to infer the main index of the *other* one. Methods of data analysis used in Machine Learning allow us to select just a few of the hundreds of variables in a database, which later are used in a panel regression, yielding a good approximation to the main index in the other database. Since the information of *SCOPUS* can be freely scraped from the Web, this approach allows to infer for free the *Impact Factor* of publications, the main index used in research assessments around the globe.

# 1 Introduction

Academic and scientific research activities have a growing economic importance, requiring larger outlays and investments both in advanced and emerging nations. In both cases it is a matter of national prestige but more than that, of strategic relevance, since the presence of large numbers of highly educated citizens contributing to the advance of human knowledge has well-established impacts on technological and industrial capabilities. These, in turn, are highly relevant to ensure the competitiveness and the economic security of nations as well as yielding other benefits to the economy (Dasgupta and David (1994) and Salter and Martin (2001)). The care and promotion of research activities should thus be a focus of public policies aimed at ensuring social and economic development (Stephan (2012) and Etzkowitz (2013)).

A particularly pressing issue in this matter is to assess the quality of the production generated by researchers in all fields of knowledge. On one hand, it is of interest to detect areas in which nationals have international impact, as to concentrate resources on them. On the other, it allows to detect weaknesses and areas and institutions requiring improvement (Abramo and D'Angelo (2014)). This is even more relevant in the case of emerging economies, in which the resources that can be devoted to these activities is rather limited (Choi (1988) and Choi and Zo (2019)).

Decision makers who have to decide on how to allocate resources in research and the promotion of academic activities need tools for making informed decisions. An important limitation is that some of the best sources of information about the quality of publications are exceedingly costly for low and low-medium income nations (Madras (2008) and Hariharan et al. (2018)) and researchers residing in these regions are publishing more frequently in the aforementioned sources (Ronda-Pupo (2021)). Thus, it is of interest for decision-makers to count with cheaper alternative sources yielding the same information.

In this paper we explore this question, analyzing how to efficiently use the limited resources available in many academic institutions to make informed decisions about the performance of researchers. More precisely, we will show how different sources of information, be they obtained freely online or via the subscription to just one of the major databases, allows to approximate the data on the performance of academic journals in which the researchers publish their work.

To proceed, we ask ourselves the following two questions:

- **Q1**. Which variables are more relevant to evaluate academic production?

- **Q2**. How should some of those variables be inferred if they are not recorded in the databases to which the authorities have access?

We answer both questions by applying methods of Econometrics and Machine Learning. More precisely, to answer **Q1** we consider the two most widely used commercial indexes of the impact of academic publications, namely *Impact Factor (IF)* (reported by Clarivate's Web of Science) and *SJR* (published in Elsevier's *SCOPUS*). We find the variables that better explain these two indicators, using *LASSO* and the *random forests* methodology.

To answer **Q2** we explore how *IF* and *SJR* can be inferred using variables drawn from *SCOPUS* and *WOS*, respectively, i.e. using the *other* database to infer each index. While the procedure is the same in both cases, we focus on ways of obtaining *IF* since *SJR* and many of its explanatory variables are already freely available online, while *IF* is only available under subscription, being thus rather inaccessible to cash challenged institutions in emerging economies.

Our main goal then is to infer the values of a given academic index using the information of the data used to determine the other index. We choose the database that might be easier to access to institutions in the developing world and apply Machine Learning methods to extract the information relevant for our purpose. This procedure would allow those institutions, which have limited funds and are not able to acquire the most costly of the two databases to make sound evaluations of the academic production of its members.

The plan of the paper is as follows. In Section 2 we discuss the problem of evaluating published research faced by academic officials in governments and in public and private institutions. We also discuss the resources available to them and the possibility of accessing them by resource limited institutions. In Section 3 we present the data and methodology we use in our study. In Section 4 we run the analyses that yield answers to **Q1** and **Q2**, and in section 5 we make explicit the policy recommendation. Finally, in section 6 we conclude.

## 2 Pros and Cons of Different Metrics

There exists a vast literature on the importance of the evaluation of the publication record of researchers for any sound science policy, adopting efficiency measures drawn from management (Stephan (2012)). This conception has been fully endorsed by emerging science powerhouses like China (Tang et al. (2020)). Some authors find that this is inherent to the scientific study of science and thus a key element for enacting optimal policies in the area (Fealing (2011)).

Among the tools for the evaluation of scientific production there are three main competitors, *IF*, *SJR* and the *h-index*. While the latter has gained popularity in the last two decades, it has been shown highly dependent on the data used to compute it (Bar-Ilan (2008)). It has even been shown that it has different versions, provided even by the same source (Jacs´o (2008) and Hu et al. (2020)).

With respect to *IF* it is the standard measure according to which publications are ranked. It is even deemed a fundamental piece in shaping research areas (Rushforth and de Rijcke (2015)). This has created incentives for its manipulation, and even the creation of "citation tribes" gaming the system (Rijcke et al. (2016)).

Be it as it may, *IF* (and increasingly also *SJR*) have become the main indicators used around the world to evaluate academic research (Harzing and Alakangas (2016)). But the high costs of subscription to *WOS* and (in a lesser extent) *SCOPUS* (Madras (2008)) have lead some authors to call for new and free versions of those indexes (Chapron and Husté (2006)).

Our take on this issue is that, given the widespread use of *IF* and *SJR*, which validates them in practice, the best policy would be to try to approximate them, when the access to them is disabled.

**3 Data and Methodology**

Our proposal is based on the analysis of the major platforms reporting data on the performance of academic journals. Additional information can be obtained by *web scraping* resources that are available online, in particular the *SCOPUS* database. It is, unlike its web interface with its search facilities, free to access. If a policy maker has access to one of the large indexation sources or to scraped information, our approach indicates how she can use the data to infer the value of the main variables in databases to which she has not access.

The statistical information needed to make this inference is obtained by merging the scraped *SCOPUS* database with that of Clarivate's *Web of Science*. This yields 21063 data entries from 1999 to 2018. We keep only the journals that are fully recorded in *both* databases (i.e. without missing values of any year nor variable). There are 3530 journals that satisfy this condition. They allow us to find the relations between the information in both databases.

The variables characterizing the quality of academic journals included in our database are listed in tables 1 and 2:

Table 1: Variables in the SCOPUS database

| Variable | Description |
|---|---|
| SJR | Scimago Journal Ranking for a given year |
| Rank | SJR Ranking position for a given year |
| Best quartile | Journal best area quartile for a given year |
| H-index | Current H-index of the journal |
| Total Docs | Total number of documents published in that year |
| Total Docs 3 years | Total docs published in the previous three years |
| Total Refs | Number of references included in the articles of a given year |
| Total Cites 3 years | Total citations received in a given year of articles published in the previous three years |
| Citable Docs 3 years | Total citable docs published in the previous three years of a given year |
| Cites/Doc 2 years | Citations received per document in a given year of articles published in the previous two years |
| Cites/Doc 3 years | Citations received per document in a given year of articles published in the previous three years |
| Cites/Doc 4 years | Citations received per document in a given year of articles published in the previous four years |
| Ref/Doc, | Average amount of references per document in a given year |
| Self cites 3 years | Total citations received by the same journal in a given year of articles published in the previous three years |
| Uncited Docs 3 years | Number of documents published in the last three years that were not cited in a given year |
| International collaboration, | Percentage of documents published produced by researchers from several countries |
| External cites 3 years | Total citations received by different journals in a given year of articles published in the previous three years |
| Non-citable docs | Total number of documents published in a given year that are not meant for citation |
| Country | Former country of the journal |
| Publisher | Publisher in charge of the journal |
| Cited Docs | Total number of documents published in a given year that are meant for citation |
| Open access | Whether the journal is available in open-access or not |
| Multidisciplinary | Whether the journal belong to a specific area of knowledge or not |
| Big areas | 26 big research areas as defined by the ASJC journal classification |
| Areas | 304 research areas as defined by the ASJC journal classification |
| Super areas | 5 big areas super groups as defined by the ASJC journal classification |

Table 2: Variables in the WoS database

| Variable | Description |
|---|---|
| IF | Journal Impact Factor for a given year |
| 5-year impact factor | Average impact factor of the last five years of a given year (included) |
| Impact factor without journal self cites | Journal Impact Factor without self citations for a given year |
| Immediacy index | Average number of times articles are cited in the year they are published for a given year |
| Citable items | Total number of documents published in a given year that are meant for citation |
| Eigenfactor score | Similar to the impact factor but weighting citations by origin for a given year |
| Article Influence score | Multiplies Eigenfactor score by 0.01 and divides by the total articles published in a given year |
| Average journal impact factor percentile | It is computed for a given year |
| Normalized eigenfactor | Normalizes the Eigenfactor score for every journal reported in a given year |

Many of these variables are categorical (for instance *publisher* or variables referring to the country of publication). We can see some of the descriptive statistics of these variables in Table 3.

As we can observe, there are also many metrics of performance. Some of those measures are objective (e.g. *number of citations* to a journal) while others combine many simpler indicator metrics. This poses the difficulty of obtaining sharp global conclusions up from such multiplicity of variables. In this kind of scenario it is usual to seek a single index summarizing the relevant information, making it more accessible to decision makers.

Table 3: Descriptive statistics

| Variable | Mean | Median | S.D. | Min. | Max. |
|---|---|---|---|---|---|
| SJR | 1.29 | 0.868 | 1.74 | 0.1 | 49.3 |
| Hindex | 90.2 | 73 | 70.8 | 0 | 1100 |
| TotalDocs | 175 | 89 | 287 | 0 | 6690 |
| TotalDocs3years | 505 | 256 | 831 | 0 | 19800 |
| TotalRefs | 5990 | 3030 | 10800 | 0 | 309000 |
| TotalCites3years | 1550 | 404 | 4860 | 0 | 130000 |
| CitableDocs3years | 467 | 242 | 779 | 0 | 19600 |
| CitesDoc2years | 2.24 | 1.6 | 2.56 | 0 | 58.3 |
| RefDoc | 39.7 | 33.9 | 31.3 | 0 | 931 |
| CitesDoc4Years | 2.47 | 1.81 | 2.76 | 0 | 60 |
| SelfCites3Years | 146 | 29 | 578 | 0 | 20500 |
| UncitedDocs3Years | 168 | 93 | 266 | 0 | 18600 |
| InternationalCollaboration | 20.4 | 19.2 | 13.2 | 0 | 100 |
| CitesDoc3Years | 2.4 | 1.75 | 2.7 | 0 | 57.9 |
| ExternalCites3Years | 1410 | 366 | 4048 | 0 | 122000 |
| NonCitableDocs | 37.9 | 7 | 139 | 0 | 4340 |
| CitedDocs | 337 | 143 | 659 | 0 | 18700 |
| JournalImpactFactor | 2.06 | 1.41 | 2.75 | 0 | 79.3 |
| ImpactFactorwithoutJournalS | 1.85 | 1.22 | 2.67 | 0 | 78.5 |
| YearImpactFactor | 2.59 | 1.92 | 3.13 | 0.01 | 70.3 |
| ImmediacyIndex | 0.428 | 0.25 | 0.66 | 0 | 20 |
| CitableItems | 157 | 80 | 268 | 0 | 6590 |
| EigenfactorScore | 0.016 | 0.00471 | 0.0566 | 0 | 1.84 |
| ArticleInfluenceScore | 0.963 | 0.636 | 1.44 | 0 | 32.6 |
| AverageJournalImpactFactorP | 56.6 | 58.6 | 26.9 | 0.159 | 99.9 |
| NormalizedEigenfactor | 1.67 | 0.518 | 5.57 | 0.00199 | 177 |
| OpenAccess | 1.04 | 1 | 0.193 | 1 | 2 |

More interestingly we can study the relation between the SCImago Journal Ranking (*SJR*) index and the Journal Impact Factor (*IF*), which are the two indexes that can be considered the representatives of the corresponding databases.

The Impact Factor of each journal is determined yearly. For instance, for a year $T$ and a journal $j$ it is defined in terms of the following variables:

- $A$: total citations in year $T$ to articles published in journal $j$ in $(T-2)$ and $(T-1)$.
- $B$: number of articles published in $j$ in $(T-2)$ and $(T-1)$.

Then, $IF(j; T) = A/B$. [1]

---
[1] https://clarivate.com/webofsciencegroup/essays/impact-factor/

The *SJR* index is obtained using the same algorithm as Google Page Rank (Chapron and Husté (2006)) and it is based on the number of references in a journal to other journals and on the number of times it has been cited in other journals. The algorithm iterates until it assigns a stationary ranking to each journal. It takes into account not only the number of cites obtained by a journal but also in which journals can they be found. Similarly for the journals that have been cited by the journal under consideration. The ranking of journals so obtained orders them according to the *eigenfactor score*( Bergstrom et al. (2008)).

In figure 1a and 1b we show two ways in which the relation between *IF* and *SJR* can be approximated. We can see that there exists a positive relation between these two indexes, but the relation is not necessarily one-to-one, with the possibility of finding journals with a high *SJR* but a low *IF* and vice versa. This is a clear evidence of the fact that there exist different alternative metrics of performance, not all of them capturing exactly the same implicit ordering of journals, and thus complicating the use of one indicator to infer the value of the other.

As an aside, the same evidence shows (see 1c) that open access journals do not achieve the highest scores in either of both indicators.

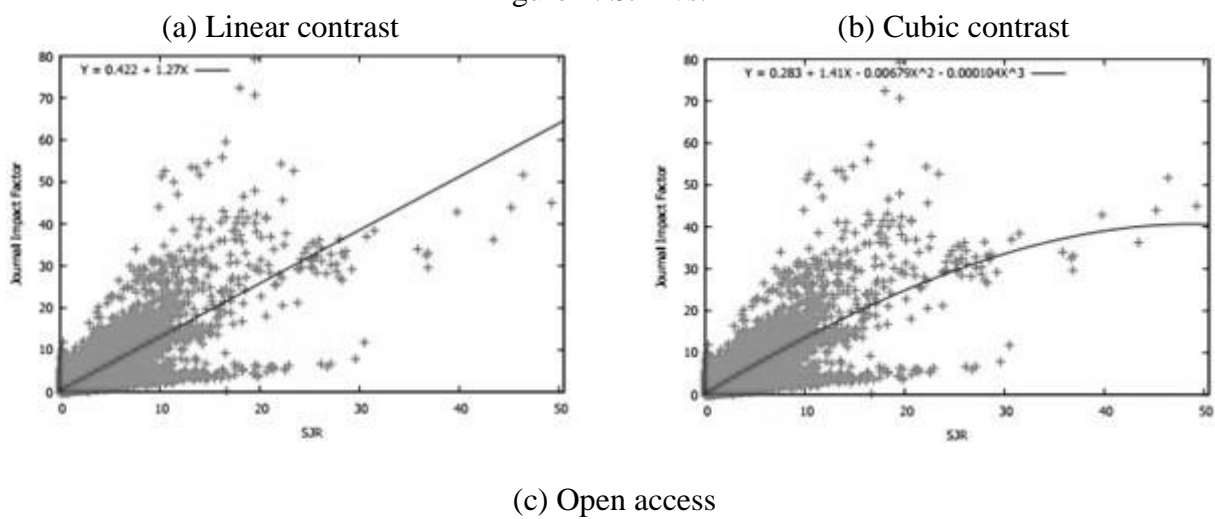

Figure 1: SJR vs. IF

(a) Linear contrast  (b) Cubic contrast

(c) Open access

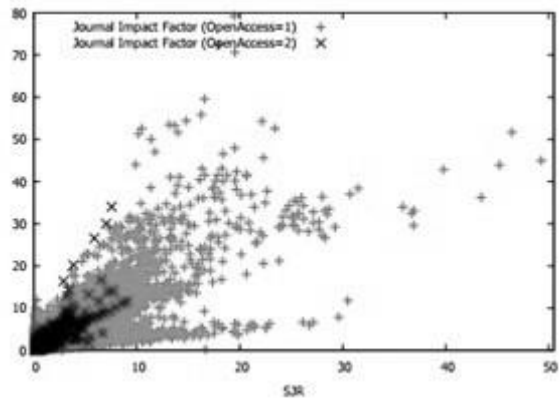

In figures 2 and 3, we can observe the relations between these indexes and some variables in their corresponding databases. At first glance, some of these variables, such as the *immediacy index* or the *article influence score* have a very positive relation with *IF* while others, such as the *eigenfactor score* do not. Interestingly, the same evidence shows no clear relation between *SJR* and relevant variables in the *SCOPUS* database.

Figure 2: SJR and some of its possible determinants

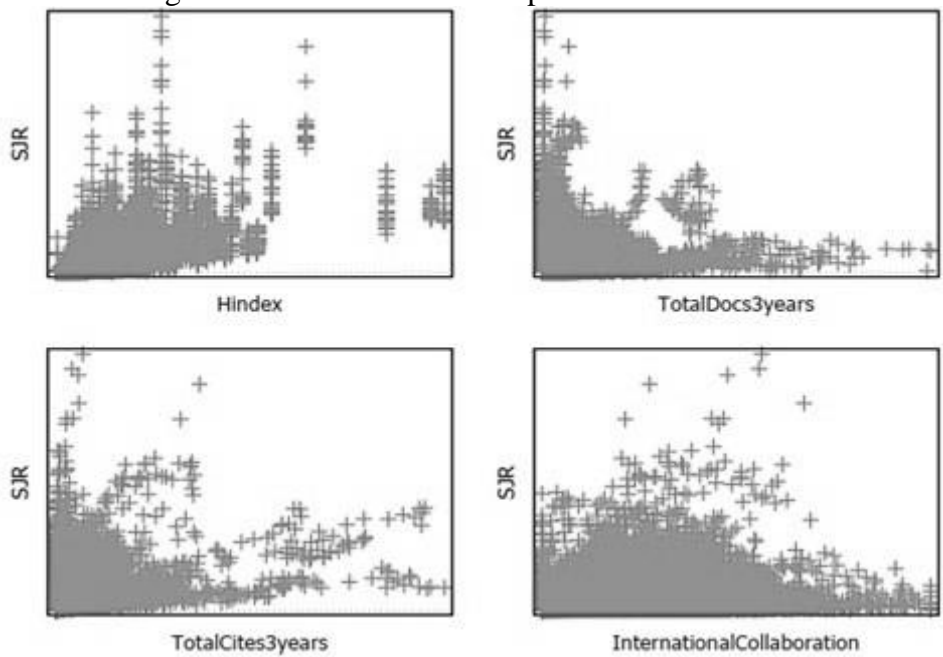

Figure 3: IF and some of its possible determinants

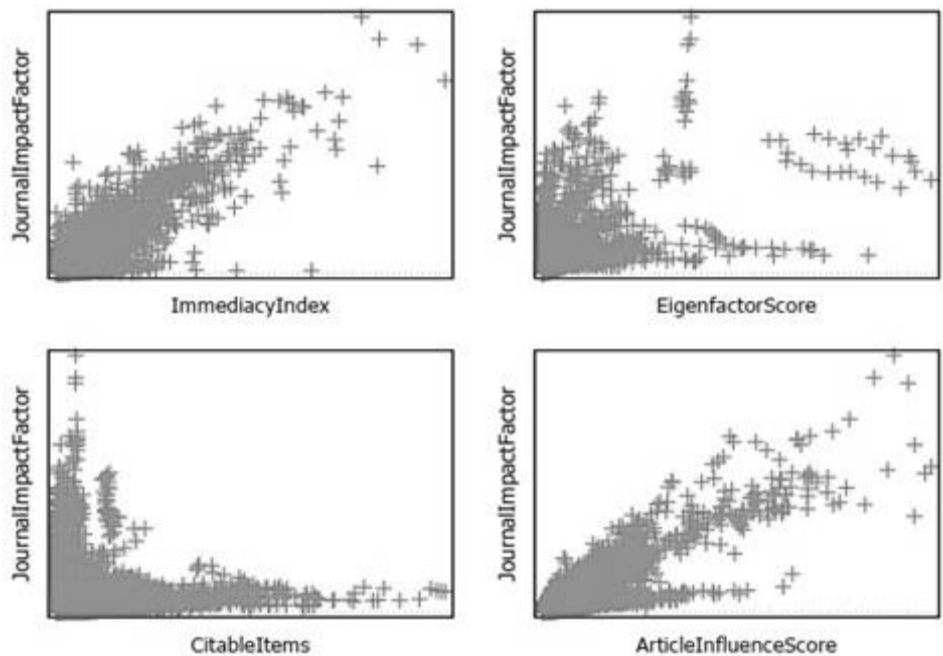

Given the aforementioned casual evidence, a first step towards providing decision makers with useful inference and prediction tools requires to find variables that are particularly relevant to infer the main indexes. Categorical ("area") variables like *scientific discipline*, *country of publication*, etc., are of interest since they may be relevant to explain the performance of journals in terms of clear political or economic reasons. On the other hand "quality" variables, indicating different aspects of that performance, may be harder to explain in plain terms but are perhaps better attuned to the intrinsic success of each journal. In both cases, methods of data analysis can be applied to select a subset of relevant variables. In our study we do this by using a combination of *LASSO* and *Random Forests* (Hastie et al. (2009)).

Once the set of relevant variables is chosen, we can run an econometric analysis of how they contribute to inferring the value of the main index variables. In our study we use a panel of journals, capturing temporal and journal fixed effects among those variables.

The coefficients obtained corresponding to the significant variables in the panel estimation are the final outcome of our analysis. Together with the data scraped from the web (or if the decision makers have a subscription to either one of the main scientometric databases), the assessment officers will be able to infer the value of the main index variables of the *other* database, to which they do not have access.

## 4 Data Exploration and Variable Selection

As said, we start with the use of data analysis techniques in order to select the variablesthat would contribute to simplify the process of making a recommendation. We discard here the possibility of using a dimension reduction procedure (as for instance considering *Principal Components*) since this involves a loss of clear interpretation of explanatory variables.

Our choice of tool involves a combination of *LASSO* and *Random Forests*. Our strategy consists in analyzing different combinations of variables:

- *SJR* against variables of *Web of Science*.
- *IF* against variables of *SCOPUS*.
- *SJR* against variables of *Web of Science* and *SCOPUS*.
- *IF* against variables of *Web of Science* and *SCOPUS*.

The goal is to select *two* groups of explanatory variables for both *SJR* and *IF*. Each of these groups will be able to explain those two main indexes. The reason for selecting two subgroups is to help to see whether the one that only includes variables to which the institution has actual access is as good as using the variables of both databases. In other words, to see whether it is possible to explain *SJR* using only features from the *WOS* dataset and explain *IF* with variables from the *SCOPUS* dataset.

In section 5 we will show the results of running regressions using "crossover" databases, executing the first and second items of the aforementioned strategy. This would simulate the situation in which decision makers have access to one database but not to the other (and want to estimate the index that is unavailable). To justify the soundness of this approach, we will investigate in the next two subsections whether the predictive power of one database to explain the other index is similar to that of running the regression against both databases.

**4.1 LASSO**

*LASSO* is a statistical method for regularization and variable selection Tibshirani (1997). It basically involves minimizing the quadratic difference between a variable and a linear expression on its explanatory variables by choosing their coefficient (i.e. an *OLS* procedure) subject to a restriction in the sum of those coefficients. If $y$ is the variable to be explained and $\{x_1, \ldots, x_n\}$ is the set of explanatory variables, and $\alpha_0, \alpha_1, \ldots, \alpha_n$ are the coefficients, the goal is, given a database $\{\langle y^i, x^i, \ldots, x_n^i \rangle\}_{i=1}^{N}$ to find

$$\sum_{i=1}^{N}\left(y^i - \alpha_0 - \sum_{j=1}^{n}\alpha_j x_j^i\right)^2$$

$$s.t. \sum_{j=1}^{n}|\alpha_j| \leq K$$

where $K$ is a given parameter of regularization. This problem can be formulated in *Lagrangian* form, where $\lambda$ is the parameter of the regularization constraint.

Notice that to each value of $\lambda$ there corresponds a value of the mean squared error (MSE). That is, for a given $\lambda$ we obtain values $\{\hat{\alpha}_j\}_{j=1}^{n}$, such that for $i = 1, \ldots, N$ we obtain $\hat{y}^i = \hat{\alpha}_0 + \sum_{j=1}^{n}\hat{\alpha}_1 x_j^i$. Then, the MSE is $\frac{1}{N}\sum_{i=1}^{N}(y^i - \hat{y}^i)^2$.

If we call $\lambda^*$ the value yielding the solution to the minimization problem, we can relax it to alternatives values $\lambda$ with a higher MSE but involving less variables (i.e. with more coefficients $\hat{\alpha}_j = 0$).

In our case, since $n > 700$, a drastic reduction in the number of variables is highly convenient. One of the main exercises we carry out is to *cross-validate* the value of $\lambda$ that ensures the minimal number of explanatory variables with the lowest possible corresponding *MSE*.

In figures 4 and 5, we depict the relation between the percentage of variability of the main index variables explained in terms of the number of variables of the *SCOPUS* and *WOS* databases allowed. That is, we depict how *MSE* varies with increasing relaxations of the value of $\lambda$.

In the (a) panels we allow $\lambda$ to relax from being 0 (i.e. not including any variable) until being large enough as to include all the variables in the database as predictors. In the (b) panels we do the same but only allowing $\lambda$ to reach a value according to which 10 explanatory features are included in the regression. In both cases we see the same pattern in which the percentage of variance explained converges quickly with the inclusion of the first predictor features and afterwards it only marginally improves with the inclusion of other predictor variables. In fact, the convergence is in general quite drastic, since the largest improvements are obtained with the first 2 or 3 variables.[2]

Figure 4: Lasso - IF vs. WOS and SCOPUS
(a) (b)

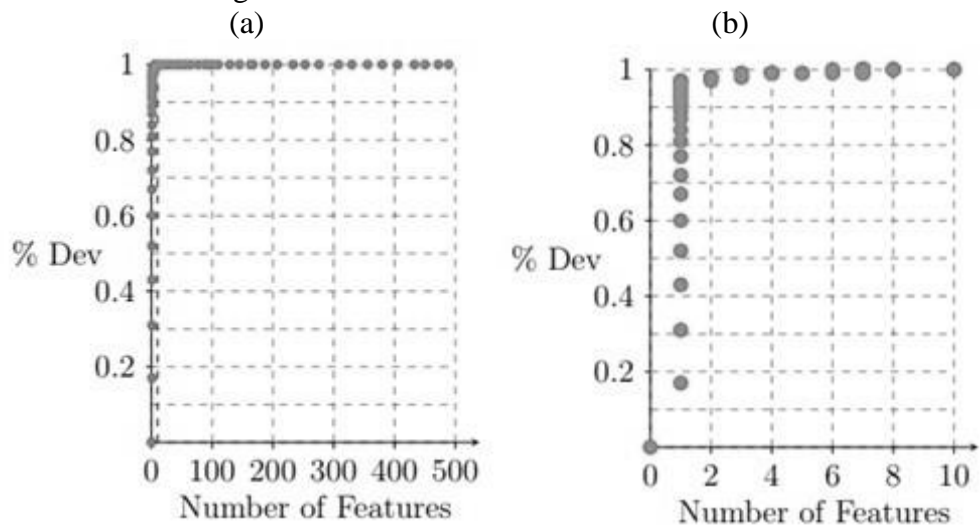

Figure 5: Lasso - SJR vs. WOS and SCOPUS
(a) (b)

---

[2] This can be related to correlation clustering as in Bansal et al. (2004), and the correlation analysis presented in the next sections of this article.

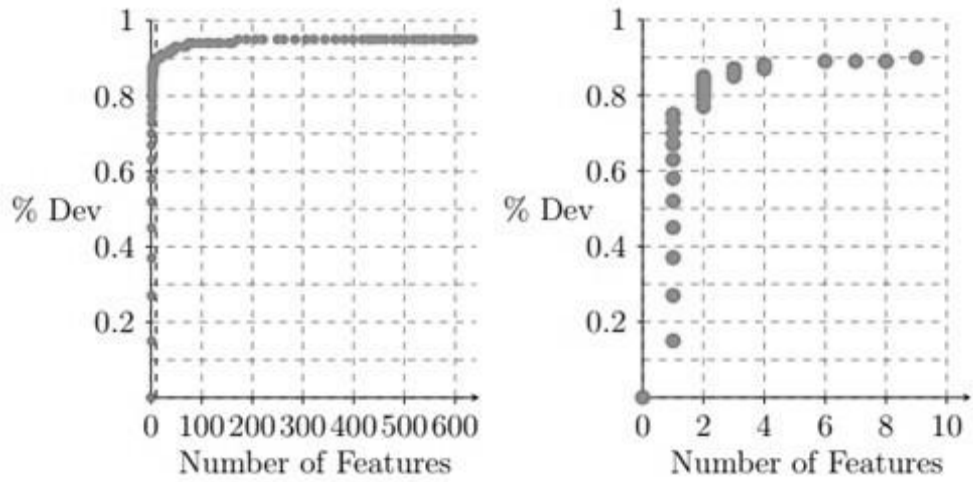

Figure 6: Lasso - SJR vs. WOS

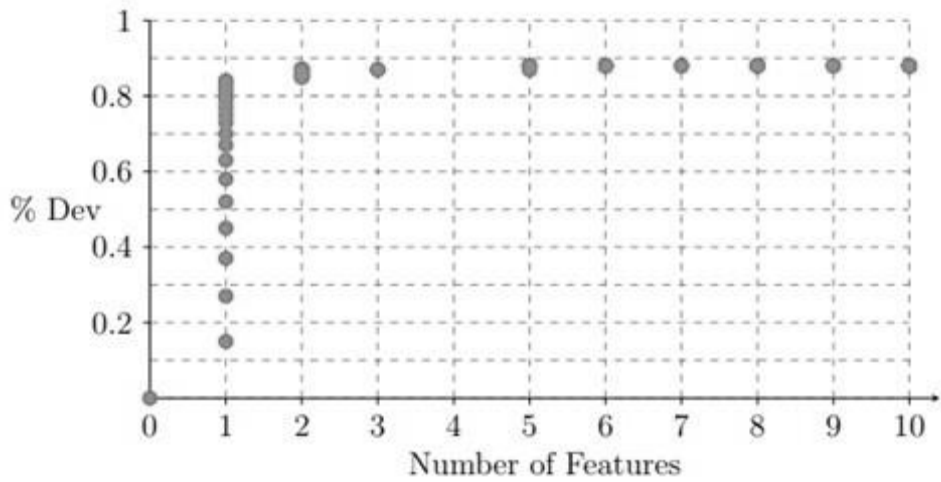

Figure 7: Lasso - IF vs. SCOPUS

(a) (b)

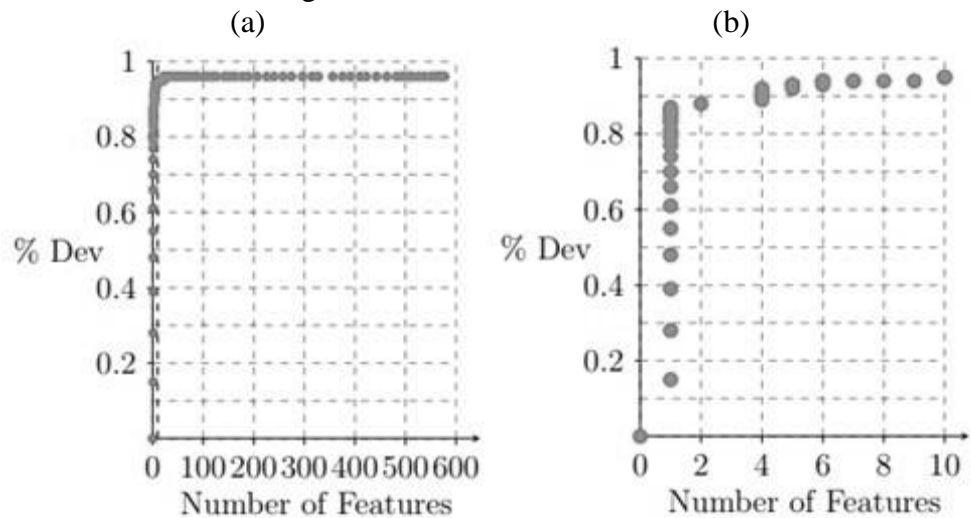

In figures 6 and 7, we repeat the analysis but only for steps 1 and 2 of the methodology. In 6 we use the first 10 predictors drawn from the *WOS* database to explain *SJR*. It can be seen that the first 4 variables explain 80% of the *MSE*. In turn, in figure 7 we fit *IF* against variables from the *SCOPUS* database.

We can see that panels (a) and (b) indicate that the results are similar to those of fitting the indexes against variables drawn from both *SCOPUS* and *WOS*. An interesting realization is that with stricter values of $\lambda$ the variables that remain are more related to quality, while the "area" ones tend to disappear. Among those we find *publisher*, *country of publication* and, perhaps more interestingly, the binary variable *open access*.

**4.2 Random Forests**

*Random forests* is a method for regression and other tasks that works by constructing a family of decision trees at training time and outputting the subclass that yields the mean prediction of the individual trees (Ho (1995)). It starts by constructing *Classification and Regression Trees* (*CART*) on training subsets drawn from the whole database under consideration. These trees are built by splitting the training set, constituting the root node of the tree, into subsets of variables - which become the successor children. The splitting is based on the quality of the prediction of some features in the training set.

The process is repeated by recursively partitioning the children sets. The recursion is completed when splitting no longer adds value to the predictions (Breiman et al. (1984)).

Then, the resulting trees are averaged out, to yield a tree in which the splitting proceeds according to the average contribution to the prediction. This reduces the variance of the resulting trees.

We run four different estimations on the entire database using *random forests*, to distinguish how variables contribute to predict some variable of interest. We order the variables according to either how the splits reduce the Mean Square Error (*MSE)* of the prediction or increase the *purity* of nodes.[3]

---
[3] The decision to split at each node is made according to a measure called *purity*. A node is totally impure when a node is split evenly and totally pure when all of its data belongs to a single class.

The figures below indicate the position of variables in the *reduction of MSE-purity* coordinate system (where *purity* is described in terms of its *Gini index*). The goal is to detect which variables yield a better classification (purity) while also minimizing the MSE. These are represented by points in the upper right part of the figure.

To highlight how the variables in our database contribute to the prediction in a *random forest* setup, we can order them in terms of their impact on the reduction of the *MSE* as well as on the *node purity index*. This is shown in figures 8 and 9. In panels (a) we show, for each variable, the contribution to the reduction of *MSE* and *node purity* of the prediction of the main index. Panels (b), as in the *LASSO* analysis, depict the same results as panels (a), but only for the ten more relevant variables. As we can see, most of the features are close to the origin of panels (a). We adopted the ad-hoc criterion of considering that the relevant contributions are those that fall in the gray areas, where both the reduction of *MSE* and *purity* are larger than 5. Finally, we represented in gray the "area" variables. It is clear that they do contribute little to both reducing *MSE* and increasing *purity*.

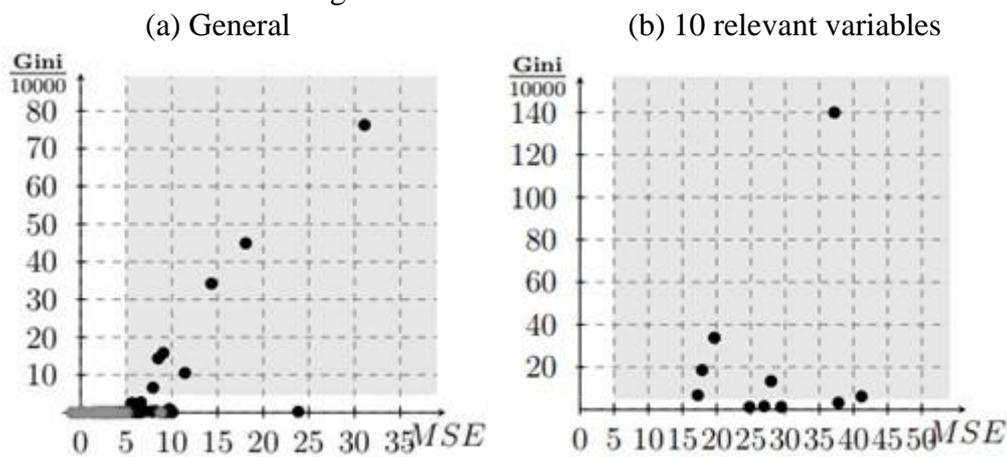

Figure 8: Random Forests – IF
(a) General           (b) 10 relevant variables

Figure 9: Random Forests – SJR
(a) General           (b) 10 relevant variables

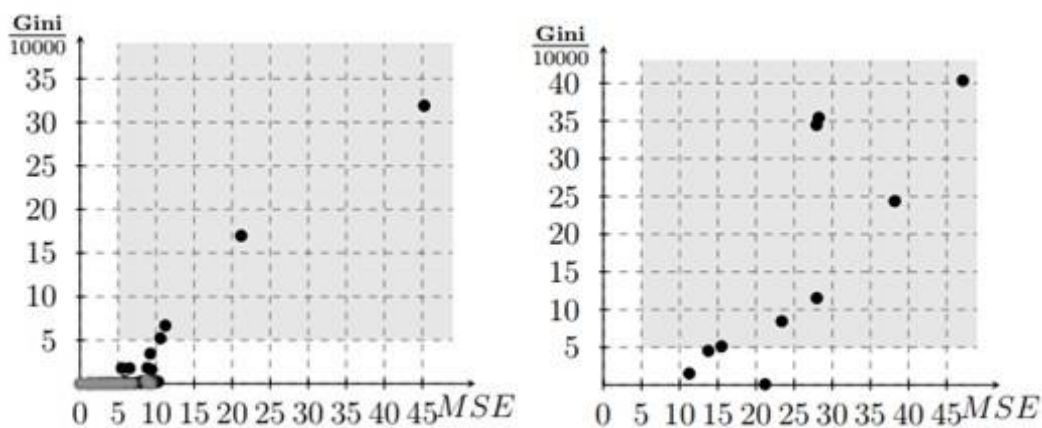

### 4.3 Selected Variables

Both *LASSO* and *random forests* indicate that "area" variables do not contribute much to the prediction of the key index variables. Instead, the top predictors are "quality" variables. Choosing them, we can reduce the complexity of the relevant model from including around 700 variables to one with just a dozen.

The selected variables are enumerated in table 4. The top of each column indicates the variable that we have chosen as being dependent in the regression that we will run, while the list of variables below are the relevant ones found by using the aforementioned methods. We use only data from the period 2013-2018, to avoid the lack of information about some variables that were not completely recorded in some years before 2013.

Notice that in table 4 the independent variables to be used in the regression of *IF* are drawn from *SCOPUS*, while those used for *SJR* are obtained from the *WOS* database.

Table 4: Variables selected for the regression analysis

| Dependent variable | |
|---|---|
| SJR | IF |
| journal impact factor (IF) | Rank |
| Eigenfactor score | SJR |
| IF without self cites | Total Docs |
| 5 year IF | Total Docs 3 years |
| Immediacy index | Total Refs |
| Citable items | Total Cites 3 years |
| Article influence score | Citable Docs 3 years |
| Average journal IF percentile | Cites/Doc 2 years |
| Normalized Eigenfactor | Ref/Doc |
|  | Cites/Doc 4 years |
|  | Self/Cites 3 years |
|  | Uncited/Docs 3 years |
|  | International Collaboration |
|  | Cites/Doc 3 years |
|  | SJR best quartile |

The idea behind this exercise is to represent cases in which a policy maker may have access to just one of the databases but intends to find a close approximation to the main index of the other major database.

Before running the regression, we check the variables to see if they are correlated. We show the results in figures 10 and 11, where we can see that most of the variables are highly (and positively) correlated. This indicates that several different variables are actually measuring basically the same phenomenon. This is true in particular for the areas surrounded by dashed lines. This procedure allows us to take care of the interdependence of the variables by choosing only one variable from each correlation cluster.

Figure 10: Correlations - SJR and relevant WOS features

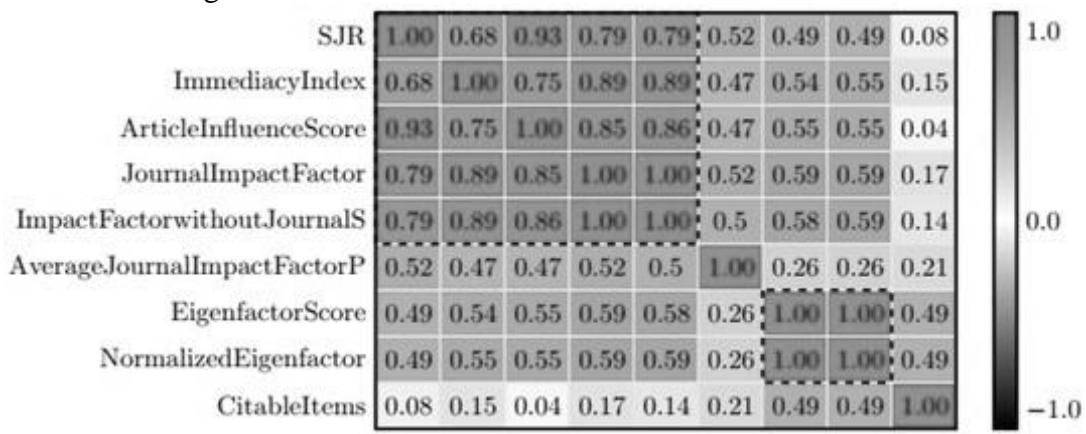

Figure 11: Correlations - IF and relevant SCOPUS features

|                          | JournalImpactFactor | CitesDocs2years | CitesDoc3Years | NonCitableDocs | ExternalCites3Years | CitedDocs | TotalDocs | TotalDocs3Years | TotalRefs | TotalCites3years | CitableDocs3years | InternationalCollaboration | RefDoc |
|--------------------------|---|---|---|---|---|---|---|---|---|---|---|---|---|
| JournalImpactFactor      | 1.00 | 0.95 | 0.94 | 0.44 | 0.59 | 0.28 | 0.27 | 0.27 | 0.3 | 0.57 | 0.21 | 0.25 | 0.38 |
| CitesDocs2years          | 0.95 | 1.00 | 0.99 | 0.32 | 0.53 | 0.28 | 0.24 | 0.24 | 0.32 | 0.52 | 0.2 | 0.3 | 0.46 |
| CitesDoc3Years           | 0.94 | 0.99 | 1.00 | 0.31 | 0.51 | 0.26 | 0.22 | 0.22 | 0.3 | 0.5 | 0.18 | 0.3 | 0.48 |
| NonCitableDocs           | 0.44 | 0.32 | 0.31 | 1.00 | 0.49 | 0.3 | 0.38 | 0.42 | 0.17 | 0.47 | 0.26 | −0.06 | −0.15 |
| ExternalCites3Years      | 0.59 | 0.53 | 0.51 | 0.49 | 1.00 | 0.78 | 0.72 | 0.76 | 0.69 | 1.00 | 0.71 | 0.14 | 0.01 |
| CitedDocs                | 0.28 | 0.28 | 0.26 | 0.3 | 0.78 | 1.00 | 0.94 | 0.97 | 0.9 | 0.81 | 0.98 | 0.13 | −0.06 |
| TotalDocs                | 0.27 | 0.24 | 0.22 | 0.38 | 0.72 | 0.94 | 1.00 | 0.96 | 0.9 | 0.76 | 0.95 | 0.08 | −0.12 |
| TotalDocs3Years          | 0.27 | 0.24 | 0.22 | 0.42 | 0.76 | 0.97 | 0.96 | 1.00 | 0.85 | 0.79 | 0.99 | 0.08 | −0.12 |
| TotalRefs                | 0.3 | 0.32 | 0.3 | 0.17 | 0.69 | 0.9 | 0.9 | 0.85 | 1.00 | 0.73 | 0.88 | 0.18 | 0.09 |
| TotalCites3years         | 0.57 | 0.52 | 0.5 | 0.47 | 1.00 | 0.81 | 0.76 | 0.79 | 0.73 | 1.00 | 0.75 | 0.14 | 0.01 |
| CitableDocs3years        | 0.21 | 0.2 | 0.18 | 0.26 | 0.71 | 0.98 | 0.95 | 0.99 | 0.88 | 0.75 | 1.00 | 0.09 | −0.1 |
| InternationalCollaboration | 0.25 | 0.3 | 0.3 | −0.06 | 0.14 | 0.13 | 0.08 | 0.08 | 0.18 | 0.14 | 0.09 | 1.00 | 0.19 |
| RefDoc                   | 0.38 | 0.46 | 0.48 | −0.15 | 0.01 | −0.06 | −0.12 | −0.12 | 0.09 | 0.01 | −0.1 | 0.19 | 1.00 |

## 5 Regressions

Once selected the relevant variables, we run a *panel data analysis*, using fixed effects (Croissant et al. (2008) and Croissant et al. (2019)). That is, once chosen a class of explanatory variables $\{x_1, \ldots, x_n\}$ and a set of variables to be explained $\{y_1, \ldots, y_m\}$, for a $t = 1, \ldots, T$ periods, the goal is to find the coefficients $\{\alpha_i\}_{i=1}^m \cup \{\alpha_j\}_{j=1}^n \{\beta_k\}_{k=1}^K$ for the following linear model:

$$y_i^t = \alpha_i + \sum_{j=1}^n \alpha_j x_j^t + \sum_{k=1}^T \beta_k \delta_k(t) + \epsilon_i^t$$

where the supraindex indicates the period of time, each $\delta_k$ is a Dirac function[4] and $\epsilon_i^t$ is an error term. Each $\alpha_i$ reflects the fixed effect on variable $y_i$.

We do not use a *dynamic panel* (in which each $y_i^t$ depends also on lagged values of $y_i$) since a panel without missing data would be rather unbalanced. This is because there are many journals with only a few observations.

---

[4] kt=1 is k=t and 0 otherwise.

We run this analysis using the R-programming language PLM package. The resulting values yield good results under both the *F test* and the *score test*, indicating that the statistical model is close to attaining the maximum *likelihood function* measuring its *goodness of fit* to the sample data. On the other hand, the *Hausman specification test* allows us to reject the possibility of yielding better results with a *random effects* model. The *F test* discards the null hypothesis that all the fixed effects are null. Therefore, we cannot simplify the analysis by running a pooled *Ordinary Least Squares* regression. Now, we consider the possibility that residuals (the difference between the actual $y_i^t$ and its predicted value according to the model) may be correlated, contradicting the assumption that each $\epsilon_i^t$ is a pure error term. To cover this case, we adjust the model using *Generalized Least Squares* (*GLS*), in which the goal is to minimize the *MSE* corrected by the *covariance matrix* (Semykina and Wooldridge (2010)).

The results under the latter model are reported in Tables 6 and 5, while those with the simpler models are in Tables 10 and 9, in the Appendix.

The results exhibit an interesting aspect of the relations between the *SCOPUS* and the *WOS* databases. While *IF* shows to be a good explanatory variable for *SJR*, neither *SJR* nor *ranking* are the most significant features explaining *IF*. It can thus be said that *WOS* explains *SCOPUS* better than the other way around. Since *WOS* has higher subscription costs, it is useful to know that, in order to predict *IF* with *SCOPUS* data (either obtained by subscription or by scraping the Web) the most relevant variables are *cites per document in the last 2* and *cites per document in the last 3 years*.

Another interesting realization is that some of the *SCOPUS* variables used to explain *IF* have negative coefficients. This is perhaps an indication that some of those variables convey the same information and thus a negative weight of some of them compensates for that redundancy.

Table 5: GLS Panel Regression for SJR

|  | Dependent variable: |
| --- | --- |
|  | SJR |
|  | (1) |
| JournalImpactFactor | 0.4158*** |
|  | -0.0246 |
| EigenfactorScore | 20.0722*** |
| | -2.4869 |
| ImpactFactorWithoutJournalSelfCites | -0.3042*** |
| | -0.0247 |
| 5YearImpactFactor | -0.0245*** |
| | -0.008 |
| ImmediacyIndex | -0.0043 |
| | -0.007 |
| CitableItems | 0.00008*** |
| | -0.00002 |
| ArticleInfluenceScore | 1.3082*** |
| | -0.0016 |
| AverageJournalImpactFactorPercentile | 0.0012*** |
| | -0.0003 |
| NormalizedEigenfactor | -0.1326*** |
| | -0.0248 |
| Observations | 21,063 |
| $R^2$ | 0.9791 |

Note: *$p<0.1$; **$p<0.05$; ***$p<0.01$

Table 6: GLS Panel Regression for IF

| | Dependent variable: |
|---|---|
| | Journal.Impact.Factor |
| | (1) |
| Rank | 0.00002 |
| | -0.00002 |
| SJR | 0.1977* |
| | -0.0082 |
| TotalDocs | 0.0002*** |
| | -0.00006 |
| TotalDocs3years | 0.0011*** |
| | -0.0001 |
| TotalRefs | -0.000004** |
| | -1E-06 |
| TotalCites3years | 0.00006*** |
| | 0 |
| CitableDocs3years | -0.0005*** |
| | -0.0001 |
| CitesDoc2years | 0.8183*** |
| | -0.018 |
| RefDoc | 0.00005 |
| | -0.00018 |
| CitesDoc4Years | -0.0608*** |
| | -0.007 |
| SelfCites3Years | -0.0001** |
| | -0.00002 |
| UncitedDocs3Years | 0.00084*** |
| | -0.0001 |
| InternationalCollaboration | 0.0002 |
| | -0.0004 |
| CitesDoc3Years | 0.035*** |
| | -0.007 |
| SJRBestQuartileQ2 | 0.011 |
| | -0.009 |
| SJRBestQuartileQ3 | 0.017 |
| | -0.017 |
| SJRBestQuartileQ4 | 0.015 |
| | -0.0323 |
| Observations | 21,063 |
| R2 | 0.986 |

Note: *p<0.1; **p<0.05; ***p<0.01

Figures 12 and 13 show the relationships between *SJR* and *IF*, respectively, with the variables with the highest coefficients in tables 12 and 13. Namely, *eigenfactor score* for *SJR*, and *cited documents* for *IF*.

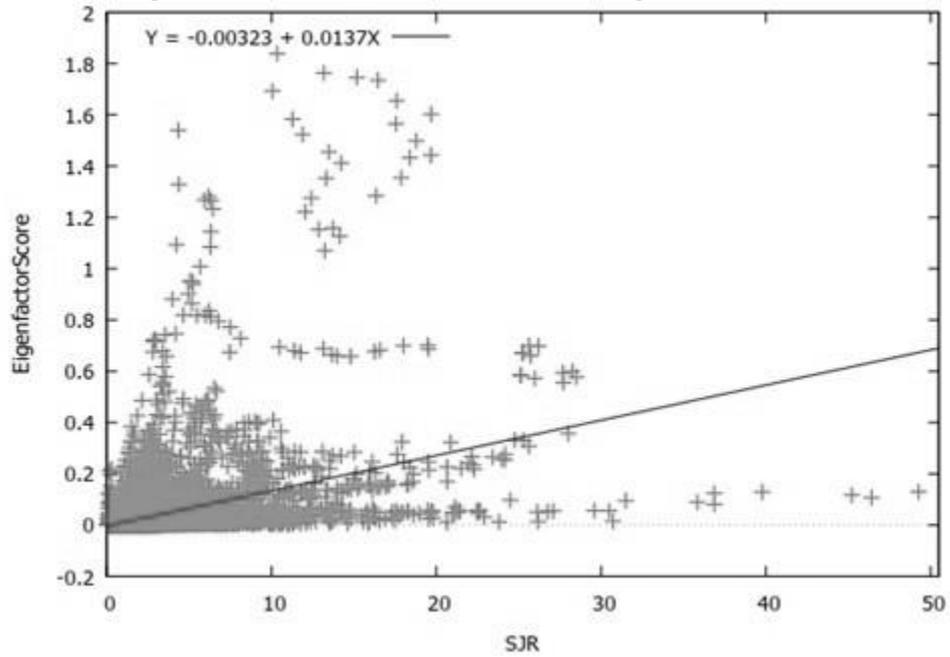

Figure 12: Relation between SJR and eigenfactor score

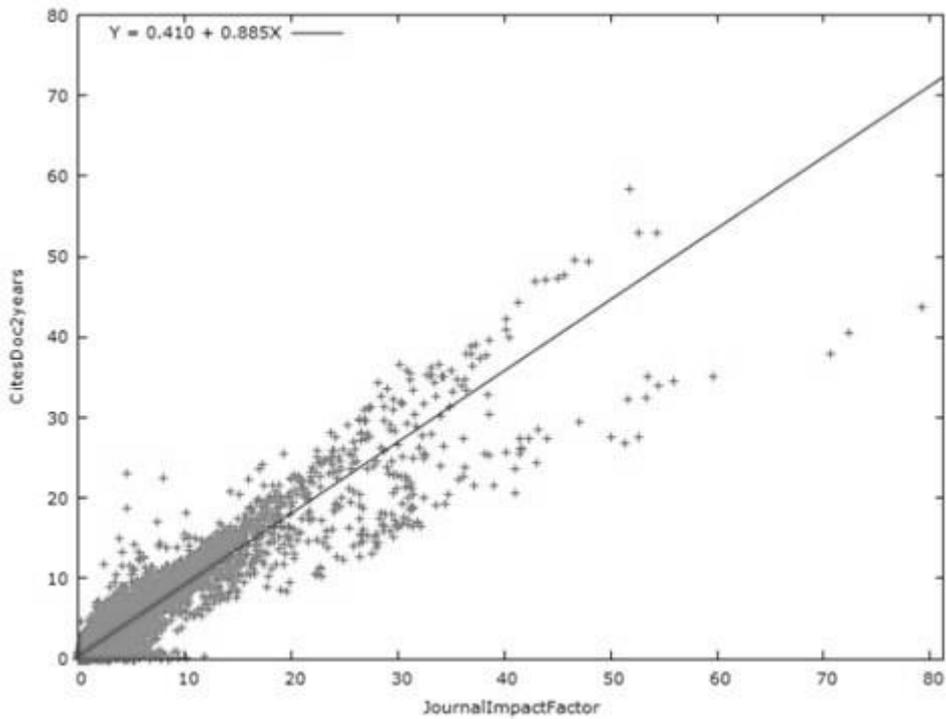

Figure 13: Relation between IF and cited documents

We have to consider the problem of multicollinearity that might be present in this case, in which variables exhibit high correlation. While this is not a very serious issue when the variables are statistically significant in spite of the increased variance, the above mentioned negative values of some coefficients may be, as said, a consequence of multicollinearity[5]. To take care of this, we run a different regression, using only variables that are representative of highly correlated clusters of variables. The results can be seen in tables 7 and 8, which show the regression coefficients for *SJR* and *IF*, respectively, with the reduced sets of variables.

These last results present a clear path towards inferring the main publication indexes using data from a different source than the actual organizations that provide the information. So, the *SJR* can be closely obtained using *IF*, *eigenscore*, *citable items* and *article influence score*. On the other hand, *IF* can be explained by *citations in documents of the last 2 years* and *international collaboration*.

The last result is not totally unexpected, since, as shown in 13, the variable corresponding to the citations in the last two years has a linear and positive relation with *IF*.

Furthermore, the formula for the calculation of *IF* includes a measure of the citations in the last biennial in its numerator. Of course, in our analysis, the value of this variable, instead of being the one reported in *WOS*, is obtained from the *SCOPUS* data.

The importance of a variable of international collaboration in explaining *IF* is, on the other hand, rather intriguing. We can only speculate that journals in which papers are authored more frequently by international teams will receive more citations since the various authors may have already presented their work to a more diverse and wide audience.

Table 7: GLS Panel Regression for SJR – II

|  | Dependent variable: |
|---|---|
|  | SJR |
|  | (1) |
| JournalImpactFactor | 0.4312*** |
|  | -0.0042 |
| EigenfactorScore | 8.3814*** |
|  | -0.4973 |
| ImmediacyIndex | -0.0038 |
|  | -0.007 |
| CitableItems | 0.00019*** |
|  | -0.00003 |
| ArticleInfluenceScore | 0.9741*** |
|  | -0.0013 |

---

[5] In fact, the results show high Variance Inflation Factors (VIF).

| | |
|---|---|
| Observations | 21,063 |
| R2 | 0.9771 |

Note: *p<0.1; **p<0.05; ***p<0.01

Table 8: GLS Panel Regression for IF – II

| | Dependent variable: |
|---|---|
| | Journal.Impact.Factor |
| | (1) |
| Rank | 0.00002 |
| | -0.00002 |
| SJR | 0.0014 |
| | -0.0072 |
| CitesDoc2years | 0.7887*** |
| | -0.0041 |
| InternationalCollaboration | 0.00387*** |
| | -0.00032 |
| TotalCites3years | 0.000009** |
| | -4E-06 |
| RefDoc | -0.00005 |
| | -0.0004 |
| SJRBestQuartileQ2 | 0.0009 |
| | -0.009 |
| SJRBestQuartileQ3 | 0.015 |
| | -0.018 |
| SJRBestQuartileQ4 | 0.021 |
| | -0.0325 |
| Observations | 21,063 |
| R2 | 0.986 |

Note: *p<0.1; **p<0.05; ***p<0.01

## 6 Policy Recommendation

The performance of researchers is usually evaluated in terms of their publications. While there is an individual aspect involved, e.g. the impact of the papers published by an author, here we focus only on the assessment of the journals in which those papers were published. Our analysis has little to offer to address the former question, but fortunately it can be answered by resorting to free-access web sites like *Google Scholar* or *Microsoft Academic*. Or by using a software tool like Harzing's *Publish or Perish*, that may access the aforementioned sites or (with a subscription) either *SCOPUS* or *WOS*. Our analysis provides a recipe to obtain a good approximation to the values of *SJR* or *IF* of journals even without having access to their corresponding sources. In particular, since *WOS* has a high cost of subscription, it becomes relevant to find a close related value for *IF*. But for this we need some variables from the *SCOPUS* database, which also requires a subscription. But unlike *WOS*, all those records can be obtained freely by scraping the Web. Thus, it is safe to assume that any academic institution has potential access to them.

While in this paper we have found the regression coefficients of a selected group of explanatory variables, in actual practice it would be recommendable to repeat the exercise, perhaps including variables that we have rejected. So for instance, the reader may have noticed that none of the variables selected is a variant of the *h-index*. The reason is that they are highly correlated with other variables in either database. But in a large-scale study they should be included. Particularly because of the well-known difference between the *h-indexes* in different disciplines (Alonso et al. (2009)).

Related to this, if different disciplines can be so distinguished, it may be also relevant to include some "area" variables, in particular those that categorize the type of topics covered or the publishing practices. This certainly will contribute to find a more accurate approximation to *IF*.

Finally, notice that in our analysis *open access* does not seem to yield sound results in either the *LASSO* or *random forests* exercises. But nothing indicates that this could not change in the future. And thus, in repetitions of the procedure, this question must be reexamined to avoid the risk of leaving relevant information aside.

**7 Conclusions**

In this work we have analyzed the variables determining the value of bibliographic performance indexes, in particular *IF* and *SJR*. The information on which this study has been carried out was obtained from the *WOS* and the *SCOPUS* databases (in the latter case also obtainable by scraping the Web). We proceeded by, first, selecting relevant variables by using Machine Learning methods. Then, with a reduced number of variables we ran a fixed effects panel regression. Afterwards, by reducing further the number of variables, keeping those that are highly correlated with other explanatory ones, we found a simple model for the approximation of the main indexes.

We prescribed this analysis as a useful tool for the evaluation of the performance of researchers and academics in budget-constrained institutions. The high costs of access to the main databases, in particular *WOS*, could hamper the possibilities of obtaining a sound evaluation. While the coefficients found in our analysis could be used right off the shelf, we recommend using a more extensive model, including some variables that we have omitted.

A natural extension of this work is to build a metric of individual performance, up from the indexes of the journals in which a researcher has published, but weighted by the citations to her papers. This information can be obtained from publicly accessible databases, without imposing an extra cost on the evaluating process.

## 8 Appendix

Table 9: Panel Regressions – SJR

|  | Dependent variable: | | | |
|---|---|---|---|---|
|  | SJR | | | |
|  | POLS | RANDOM | FIXED | FIXED & TIME |
| JournalImpactFactor | 0.577*** | 0.456*** | 0.457*** | 0.575*** |
|  | -0.022 | -0.025 | -0.022 | -0.022 |
| EigenfactorScore | 41.496*** | 24.657*** | 37.169*** | 39.129*** |
|  | -3.504 | -1.995 | -1.776 | -3.639 |
| ImpactFactorWithoutJournalSelfCites | -0.579*** | -0.306*** | -0.333*** | -0.577*** |
|  | -0.021 | -0.025 | -0.022 | -0.021 |
| 5.YearImpactFactor | -0.021*** | -0.274*** | -0.202*** | -0.020*** |
|  | -0.008 | -0.008 | -0.007 | -0.008 |
| ImmediacyIndex | -0.151*** | -0.029*** | -0.048*** | -0.148*** |
|  | -0.012 | -0.008 | -0.008 | -0.012 |
| CitableItems | 0.0002*** | -0.00004 | 0.0001*** | 0.0002*** |
|  | -0.00002 | -0.00003 | -0.00002 | -0.00002 |
| ArticleInfluenceScore | 1.234*** | 1.380*** | 1.321*** | 1.233*** |
|  | -0.007 | -0.015 | -0.011 | -0.007 |
| AverageJournalImpactFactorPercentile | 0.005*** | 0.002*** | 0.003*** | 0.005*** |
|  | -0.0002 | -0.0003 | -0.0003 | -0.0002 |
| NormalizedEigenfactor | -0.384*** | -0.176*** | -0.334*** | -0.363*** |
|  | -0.031 | -0.02 | -0.016 | -0.032 |
| Constant | -0.075*** |  | 0.099*** |  |
|  | -0.011 |  | -0.017 |  |
| Observations | 21,063 | 21,063 | 21,063 | 21,063 |
| R2 | 0.883 | 0.456 | 0.684 | 0.883 |
| Adjusted R2 | 0.883 | 0.346 | 0.684 | 0.883 |
| F Statistic | 17,667.120*** (df = 9; 21053) | 1,632.257*** (df = 9; 17527) | 45,505.710*** | 17,661.440*** (df = 9; 21048) |

Note: *p<0.1; **p<0.05; ***p<0.01

Table 10: Panel Regressions – IF

|  | Dependent variable: | | | |
|---|---|---|---|---|
|  | Journal.Impact.Factor | | | |
|  | POLS | RANDOM | FIXED | FIXED & TIME |
| Rank | 0.00002*** | 0.00002*** | 0.00002*** | 0.00001*** |
|  | 0 | 0 | 0 | 0 |
| SJR | 0.109*** | -0.071*** | 0.007 | 0.110*** |
| | -0.005 | -0.01 | -0.008 | -0.005 |
| Hindex | 0.002*** | | 0.004*** | 0.002*** |
| | -0.0002 | | -0.0003 | -0.0002 |
| TotalDocs | 0.001*** | -0.00003 | 0.0002*** | 0.001*** |
| | -0.0001 | -0.0001 | -0.0001 | -0.0001 |
| TotalDocs3years | 0.0003*** | -0.001*** | 0.00003 | 0.0002*** |
| | -0.0001 | -0.0001 | -0.0001 | -0.0001 |
| TotalRefs | -0.00000** | 0.00002*** | 0.00001*** | -0.00001*** |
| | 0 | 0 | 0 | 0 |
| TotalCites3years | 0.0001*** | 0.0001*** | 0.0001*** | 0.0001*** |
| | 0 | -0.00001 | 0 | 0 |
| CitableDocs3years | -0.001*** | 0 | -0.001*** | -0.001*** |
| | -0.0001 | -0.0001 | -0.0001 | -0.0001 |
| CitesDoc2years | 1.136*** | 0.868*** | 0.900*** | 1.126*** |
| | -0.013 | -0.008 | -0.008 | -0.013 |
| RefDoc | 0.002*** | -0.0001 | 0.001*** | 0.002*** |
| | -0.0002 | -0.0003 | -0.0002 | -0.0002 |
| CitesDoc4Years | -0.315*** | -0.069*** | -0.064*** | -0.298*** |
| | -0.016 | -0.011 | -0.01 | -0.015 |
| SelfCites3Years | -0.0001*** | -0.0003*** | -0.0003*** | -0.0001*** |
| | -0.00002 | -0.00004 | -0.00002 | -0.00002 |
| UncitedDocs.3.Years | 0.001*** | 0.001*** | 0.002*** | 0.002*** |
| | -0.0001 | -0.0001 | -0.0001 | -0.0001 |
| InternationalCollaboration | 0.0002 | 0.008*** | 0.006*** | -0.001*** |
| | -0.0004 | -0.001 | -0.0005 | -0.0004 |
| CitesDoc3Years | 0.074*** | 0.066*** | 0.040*** | 0.066*** |
| | -0.021 | -0.013 | -0.012 | -0.021 |
| SJRBestQuartileQ2 | 0.077*** | -0.017 | 0.002 | 0.086*** |
| | -0.016 | -0.015 | -0.014 | -0.015 |
| SJRBestQuartileQ3 | 0.032 | -0.047* | -0.03 | 0.058** |
| | -0.027 | -0.028 | -0.026 | -0.027 |
| SJRBestQuartileQ4 | 0.042 | -0.072 | -0.054 | 0.088* |
| | -0.053 | -0.052 | -0.049 | -0.052 |
| Constant | -0.518*** | | -0.613*** | |
| | -0.025 | | -0.033 | |
| Observations | 21,063 | 21,063 | 21,063 | 21,063 |
| R2 | 0.943 | 0.64 | 0.834 | 0.945 |
| Adjusted R2 | 0.943 | 0.567 | 0.834 | 0.945 |
| F Statistic | 19,227.450*** (df = 18; 21044) | 1,828.156*** (df = 17; 17519) | 106,021.200*** | 20,076.030*** (df = 18; 21039) |

Note: $^*p<0.1$; $^{**}p<0.05$; $^{***}p<0.01$